# Zero to 16383 Through the Wire: Transmitting High-Resolution MIDI with WebSockets and the Browser

**Daniel McKemie**
Independent Artist
info@danielmckemie.com

## ABSTRACT

*This paper outlines how to leverage the Web MIDI API and web technologies to convert numerical data in JavaScript to Most Significant Byte and Least Significant Byte combos, stage the data as dual concurrent CC messages, use WebSockets to send it to multiple endpoints, and wire the browser to other music software. This method allows users to control their own native application via 14-bit MIDI messaging and even applications housed on a remote source. Because the technology utilizes WebSockets, it is not reliant on local networks for connectivity and opens the possibilities of remote software control and collaboration anywhere in the world. While no shortage of options exists for controlling music software from the web, the Web MIDI API allows for a more streamlined end user experience as it seamlessly links to core OS MIDI functionality. The paper will share a use case of transmitting high-resolution MIDI through the browser and translating it to control voltage data for use with a modular synthesizer.*

## INTRODUCTION

MIDI was introduced in the 1980s and was purposefully left open ended in some places to allow for future developments and maximum flexibility for hardware manufacturers adopting it in their products [1]. The Web MIDI API is a set of interfaces that deal with the practical aspects of sending and receiving MIDI messages in the browser. It opens the web browser to direct connectivity with any software or hardware device equipped with MIDI.

## 14-BIT MIDI AND THE WEB MIDI API

Non-registered Parameter Numbers, or NRPNs, combine MIDI messages to give more granular and continuous control [2]. Because of the issues around the 31250 baud rate and the size of NRPN messages compared to normal messages, 9 bytes and 3 bytes, respectively; many hardware manufacturers were not quick to adopt this for wider use in their products as it could cause playback and synchronization issues [3].

With software allowing for any number of data types to be employed, the use case for Web MIDI discussed here is the seamless integration of the browser to native musical applications and connecting them through the internet.

While using Web MIDI to connect native applications over the internet is not novel [4, 5], the focus of this paper is on the packaging, transfer, and programmatic treatment of 14-bit MIDI messages across the internet, utilizing the Document Object Model for end user interaction and control of the data, and using high-resolution messages in various ways, specifically here as control voltage for a modular synthesizer as in my piece *Decontrol*, to be discussed later.

## WEBSOCKETS AND TRANSFER

### 1.1 Capturing and Packaging the Data on the Client

The Web MIDI API handles the client-side operations and data packaging in the browser and the WebSocket API facilitates the transfer of the packaged data over the internet. Node.js and the default Socket.io parser are both employed to handle the server and transfer[1].

```
function sendMessage(x, y, a, chan) {
        let val1 = a & 127;
        let val2 = a >> 7;
        socket.emit(outputSocket, {
                msbx: x,
                msby: y,
                lsbx: val2,
                lsby: val1,
                channel: chan
        });
}
```

---

[1]    https://www.socket.io

Figure 1. The function of packaging and sending the MSB/LSB combo over a WebSocket.

## 1.2 Transferring and Output on the Receiving Side

The packaging and transmission of the NRPN message is defined through a JavaScript function that has four parameters:

- **x** = first CC#

- **y** = second CC#

- **a** = the value 0-16383 provided by the DOM GUI element on the web page

- **chan** = the MIDI channel in which the message will be transferred

Figure 2. Four parameters of the *sendMessage* function.

The same two control channels are used for the most significant, MSB CC# 38; and least significant byte, LSB CC #06. In the *sendMessage* function, the single value of *a* is passed into two variables with bitwise operators translating the values and assigning them to the respective MSB/LSB attributes in the object prior to transmission. The channel assignment functions as normal to segment and route the MIDI data at the user's discretion at the receiving end, in this case giving each DOM element its own respective channel. The NRPN data is captured on the other end and output directly in its own function and sent to Max/MSP for further treatment and translation into a control voltage signal for a modular synthesizer.

```
socket.on('midiTransport-1', function(data) {
    output.setNonRegisteredParameter([data.msbx,
data.msby], [data.lsbx, data.lsby], data.chan-
nel);
});
```

Figure 3. Capture and output of the NRPN data to Max/MSP.

# USE CASE: DECONTROL

## 2.1 Document Object Model

*Decontrol* for live electronics and audience uses this architecture with Max/MSP to convert 14-bit MIDI values into control voltage signals for a modular synthesizer. The piece invites the audience to visit a provided URL to a web page equipped with HTML sliders and button. The elements fade in and out on the page, which are timed differently for each user visiting the page, creating a gamified environment.

```
allParams.frequency.addEventListener('input',
function() {sendMessage(38, 06, allParams.fre-
quency.value, 1);
});
```

Figure 4. Using the Document Object Model to link an HTML slider to the NRPN message function.

## 2.2 Max/MSP

Using Max/MSP as the intermediary software between the browser and the hardware, the NRPN MIDI message is first sent to Max via the browser.

The values of CC #38 and CC #6 are grabbed from the browser and passed as two arguments into an expression object that performs the reverse bitwise operation from the JavaScript function discussed earlier. This integer is scaled to an exponential floating-point curve of 0-0.9 and converted to a direct current signal patched to the modular synthesizer.

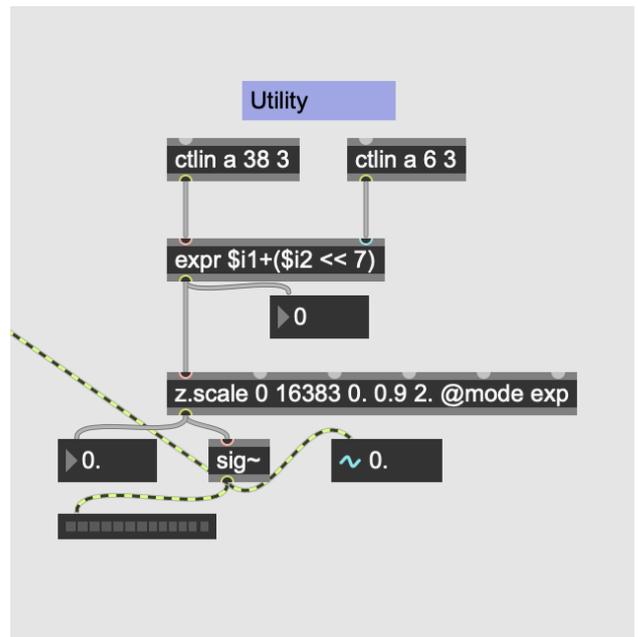

Figure 3. Max patch excerpt to convert MIDI data to integer to signal date, and output to a modular synthesizer.

On the receiving side, the patch point descriptions are left intentionally vague to fit any system, give the performer flexibility on musical choices for patching, and not bog the audience down with technical jargon or expectations on the exact behavior they are seeking to change.

# LIMITATIONS

There are countless combinations of instrument designs and collaborative works that can be developed using web technologies; however, there stands an issue with the Web MIDI API in that it is not an entirely secure browser feature. As of the date of this writing, neither Safari nor Firefox support Web MIDI, though the latter does with a dedicated extension. Google Chrome has full support, but the security of the API remains in question[2].

With Max/MSP being the example native applicated for this use case, there exists other possibilities for wiring Max to the web browser, most notably using Node directly. The level of customization is not so afforded with all music software, especially when compared to MIDI. It is the hope of the author that further development on a more secure API is in the future, with at least granting the browser the ability to send and receive CC messages without the need for requiring any sort of serial or Sysex communication. Regardless of this future state, it remains likely that employing an expanded server-side integration using Node for Max would serve the same function. But would then increase the dependency of Max, which may not suit all end users.

## FUTURE WORK

As stated above, it is the goal to tighten this architecture to be as modular as possible and bypass the concerns of standing security concerns. This includes tasks such as solutions for wider browser support, more effective middleware, and plug and play hardware.

Implementing Arduino or Daisy Seed boards for a completely integrated solution is in the works; with a sound source and native DAC to eliminate the need for any outboard modular synthesizer, or the reliance on a DC-coupled interface. While contributing to the Web MIDI API itself is not a feasible path, what is, is contributing to a tighter model of the peripheral components. A use case of incorporating Web Audio alongside this architecture will find a solution completely independent from hardware. A future performance of this piece is in the works that seeks to eliminate the modular synthesizer

There remains an openness to the piece for further manipulation of sound and visual components with this MIDI data. Projecting the visual components of the piece could increase audience engagement or become part of a score [6].

Integrating audio and video alongside high-def MIDI is currently in the works, using Web Audio, WebRTC and UDP to facilitate a faster data exchange [7]. It is the goal to have all these components tied together for a lightweight plug and play solution that is not dependent on any external downloads, operating system, or hardware.

### Acknowledgments

Many thanks to Nick Wang for opening my eyes to the depths of the MIDI protocol. And to Hank Mason for his outstanding creative use of it on his record *Memory Buffer*.